\documentstyle[sprocl]{article}

\input{psfig}

\def\be{\begin{equation}}
\def\ee{\end{equation}}
\def\bea{\begin{eqnarray}}
\def\eea{\end{eqnarray}}

\makeatletter
\newbox\slashbox \setbox\slashbox=\hbox{\large$/$}
\def\pslash#1{\setbox\@tempboxa=\hbox{$#1$}
  \@tempdima=0.5\wd\slashbox \advance\@tempdima 0.5\wd\@tempboxa
  \copy\slashbox \kern-\@tempdima \box\@tempboxa}

\def\@cite#1#2{{#1\if@tempswa , #2\fi}}
\def\@citex[#1]#2{%
\if@filesw\immediate\write\@auxout{\string\citation{#2}}\fi
\leavevmode\unskip$^{\,\scriptstyle\@cite{\@collapse{#2}}{#1}}$}
\def\CITE{\@ifnextchar[{\@tempswatrue\@CITEX}{\@tempswafalse\@CITEX[]}}
\let\onlinecite\CITE
\def\@CITEX[#1]#2{%
\if@filesw\immediate\write\@auxout{\string\citation{#2}}\fi
\leavevmode\unskip\ \@cite{\@collapse{#2}}{#1}}
\def\@collapse#1{{%
\let\@temp\relax
\@tempcntb\@MM
\def\@citea{}%
\@for \@citeb:=#1\do{%
\@ifundefined{b@\@citeb}%
{\@temp\@citea{\bf ?}%
\@tempcntb\@MM\let\@temp\relax
\@warning{Citation `\@citeb ' on page \thepage\space undefined}}%
{\@tempcnta\@tempcntb \advance\@tempcnta\@ne
\edef\MyTemp{\csname b@\@citeb\endcsname}%
\def\@tempa{\@temptokena=\bgroup}%
\afterassignment\@tempa %
\@tempcntb=0\MyTemp\relax}%
\ifnum\@tempcntb=0\relax%
\@tempcntb=\@MM
\@citea\MyTemp
\let\@temp = \relax
\else %
\edef\@tempd{\number\@tempcntb}%
\ifnum\@tempcnta=\@tempcntb %
\ifx\@temp\relax %
\edef\@temp{\@citea\@tempd}%
\else
\edef\@temp{\hbox{--}\@tempd}%
\fi
\else %
\@temp\@citea\@tempd
\let\@temp\relax
\fi
\fi
}%
\def\@citea{, }%
}%
\@temp %
}}%

\makeatother

\begin{document}

\title{EIGENVALUES OF THE QCD DIRAC OPERATOR AT FINITE TEMPERATURE AND DENSITY}

\author{E. BITTNER$^{\rm a}$, M.-P. LOMBARDO$^{\rm b}$, H. MARKUM$^{\rm a}$,
        R. PULLIRSCH$^{\rm a}$}
\address{$^{\rm a}$Institut f\"ur Kernphysik, Technische Universit\"at Wien, \\
                A-1040 Vienna, Austria\\
$^{\rm b}$Istituto Nazionale di Fisica Nucleare, \\
         Sezione di Padova, e Gruppo Collegato di Trento, Italy}

\maketitle

\abstracts{
We investigate the eigenvalue spectrum of the staggered Dirac matrix 
in two-color QCD at nonzero temperature and at baryon density when the 
eigenvalues become complex. The quasi-zero modes and their role for 
chiral symmetry breaking and the deconfinement transition are examined. 
The bulk of the spectrum and its relation to quantum chaos is considered. 
Comparison with predictions from random matrix theory is presented. 
} 

\sloppy

\section{Chiral Condensate}

The properties of the eigenvalues of the Dirac operator are of great
interest for important features of QCD. The
Banks-Casher formula \cite{Bank80} relates the Dirac eigenvalue density
$\rho(\lambda)$ at $\lambda=0$ to the chiral condensate,
$ \Sigma \equiv |\langle \bar{\psi} \psi \rangle| =
 \lim_{\varepsilon\to 0}\lim_{V\to\infty} \pi\rho (\varepsilon)/V$.
The microscopic spectral density, $ \rho_s (z) = \lim_{V\to\infty} 
 \rho \left( {z/V\Sigma } \right)/V\Sigma , $
should be given by the appropriate result of random matrix theory 
(RMT),~\cite{ShVe92} which
also generates the Leutwyler-Smilga sum rules.~\cite{LeSm92}

A formulation of the QCD Dirac operator at chemical potential
$\mu\ne0$ on the lattice in the staggered scheme is
given by \cite{Hase83}
\begin{eqnarray}
  \label{Dirac}
  M_{x,y}(U,\mu) & = &
  \frac{1}{2a} \sum\limits_{\nu=\hat{x},\hat{y},\hat{z}}
  \left[U_{\nu}(x)\eta_{\nu}(x)\delta_{y,x\!+\!\nu}-{\rm h.c.}\right]
  \nonumber\\
  & + &
   \frac{1}{2a}\left[U_{\hat{t}}(x)\eta_{\hat{t}}(x)e^{\mu}
    \delta_{y,x\!+\!\hat{t}}
    -U_{\hat{t}}^{\dagger}(y)\eta_{\hat{t}}(y)
    e^{-\mu}\delta_{y,x\!-\!\hat{t}}\right] \: ,
\end{eqnarray}
with the link variables $U$ and the staggered phases $\eta$.
We report on computations with gauge group SU(2) on a $6^4$ lattice
at $\beta=4/g^2=1.3$ and with $N_f=2$ flavors
of staggered fermions of mass $m=0.07$. For this system the fermion
determinant is real and lattice simulations of the full theory with
chemical potential become feasible exhibiting a  phase
transition at $\mu_c \approx m_{\pi}/2 \approx 0.3$ where the chiral condensate 
(nearly) vanishes and a diquark condensate develops.~\cite{Hand99}

\begin{figure}
   \begin{center}
     \psfig{figure=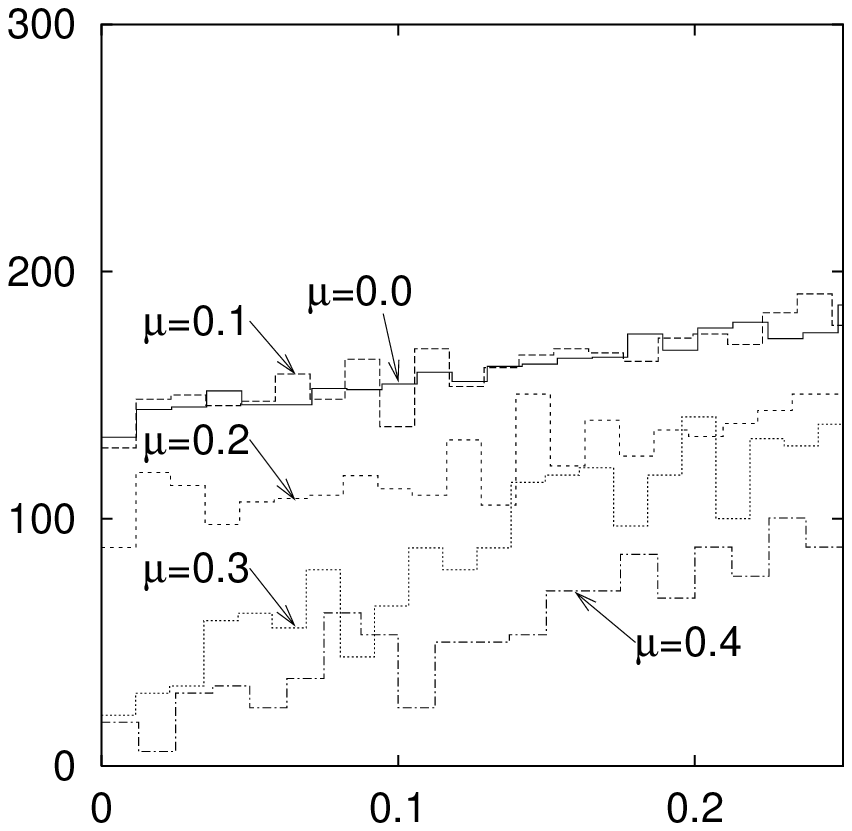,width=4cm}\hspace*{15mm}
     \psfig{figure=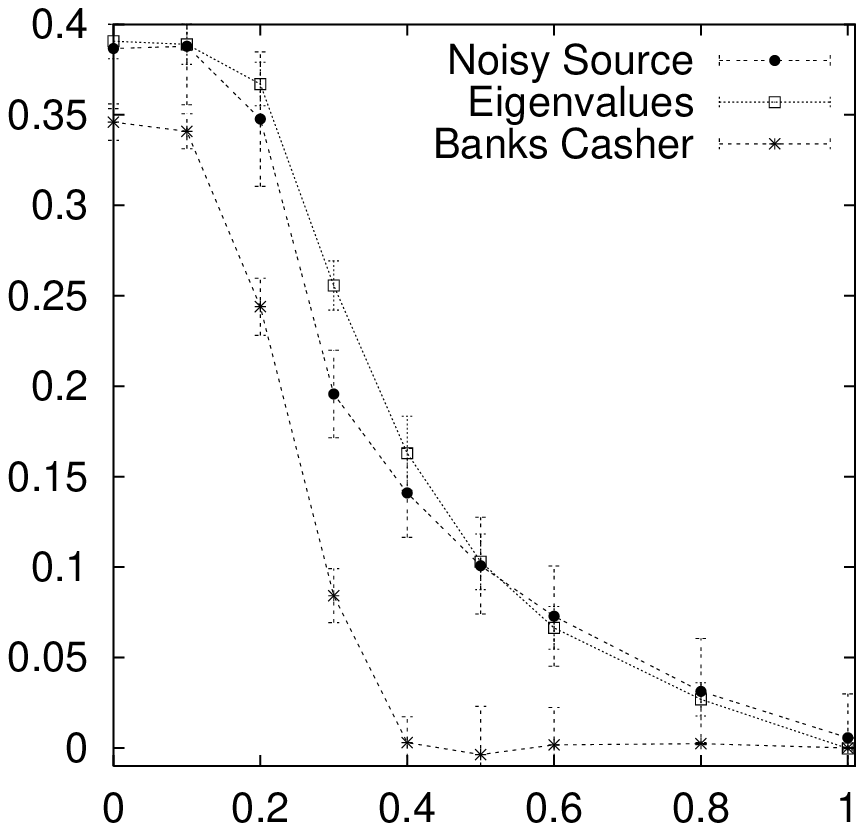,width=4cm}\\[-40mm]
     \hspace*{-50mm}$\rho(y)$\hspace*{50mm}$\langle\bar\psi\psi\rangle$\\[32mm]
     \hspace*{39mm}$y$\hspace*{61mm}$\mu$
   \end{center}
   \caption{Left plot: Density $\rho(y)$ of small eigenvalues
        for two-color QCD on a $6^4$ lattice from $\mu=0$ to 0.4.
        The loss of quasi-zero modes is accompanied by a vanishing
        of the chiral condensate.
            Right plot: Chiral condensate extracted by three different
        methods (see text).
       }
   \label{fig1}
 \end{figure}

In the left plot of Fig.~\ref{fig1} we compare the
densities of the small eigenvalues at $\mu=0$ to 0.4 on our
$6^4$ lattice, averaged over at least 160 configurations. Since the eigenvalues
move into the complex plane for $\mu > 0$, a band of width $\epsilon
= 0.015$ parallel to the imaginary axis is considered to construct
$\rho(y)$, i.e. $\rho(y)\equiv\int_{-\epsilon}^\epsilon
dx\,\rho(x,y)$, where $\rho(x,y)$ is the density of the complex
eigenvalues $x+iy$.

The density $\rho(y)$ is used to estimate a value for the chiral condensate
by naively applying the Banks-Casher relation which originally was derived
for real eigenvalues appearing in pairs of opposite sign. We further
employed the standard definition of the Green's function~\cite{LeSm92} by
inverting the fermionic matrix with a noisy source and by computing its
eigenvalues exactly, respectively, to get
the condensate. Thus the chiral condensate for two-color QCD with
finite chemical potential was extracted by three methods. The
preliminary results for $<\bar \psi \psi>$ and its variance are shown
in the righthand plot of Fig.~\ref{fig1}.

\section{Quantum Chaos}

The fluctuation properties of the eigenvalues in the bulk of the
spectrum have also attracted attention. It was shown in
Ref.~\onlinecite{Hala95} for Hermitian Dirac operators that on
the scale of the mean level spacing they are described by RMT.
For example, the nearest-neighbor spacing
distribution $P(s)$, i.e. the distribution of spacings $s$ between
adjacent eigenvalues on the unfolded scale, agrees with the Wigner
surmise of RMT.  According to the Bohigas-Giannoni-Schmit
conjecture,~\cite{Bohi84} quantum systems whose classical counterparts are
chaotic have a nearest-neighbor spacing distribution given by RMT
whereas systems whose classical counterparts are integrable obey a
Poisson distribution, $P_{\rm P}(s)=e^{-s}$.  Therefore, the specific
form of $P(s)$ is often taken as a criterion for the presence or
absence of ``quantum chaos''.

For $\mu>0$, the Dirac operator
loses its Hermiticity properties so that its eigenvalues become
complex. The aim of the present analysis is to investigate whether
non-Hermitian RMT is able to describe the fluctuation properties of
the complex eigenvalues of the QCD Dirac operator. We apply a
two-dimensional unfolding procedure~\cite{Mark99} to separate the
average eigenvalue 
density from the fluctuations and construct the nearest-neighbor
spacing distribution, $P(s)$, of adjacent eigenvalues in the complex
plane. Adjacent eigenvalues are defined to be the pairs for which the
Euclidean distance in the complex plane is smallest.  
The data are then compared to analytical predictions of the Ginibre
ensemble \cite{Gini65} of non-Hermitian RMT, which describes the
situation where the real and imaginary parts of the strongly
correlated eigenvalues have approximately the same average magnitude.
In the Ginibre ensemble, the average spectral density is already
constant inside a circle and zero outside.  In this
case, unfolding is not necessary, and $P(s)$ is given by \cite{Grob88}
\begin{equation} \label{Ginibre}
  P_{\rm G}(s)  =  c\, p(cs)\:, ~~p(s) = 
  2s\lim_{N\to\infty}\left[\prod_{n=1}^{N-1}e_n(s^2)\,e^{-s^2}
  \right] \sum_{n=1}^{N-1}\frac{s^{2n}}{n!e_n(s^2)}\:,
\end{equation}
where $e_n(x)=\sum_{m=0}^n x^m/m!$ and $c=\int_0^\infty ds \, s \,
p(s)=1.1429...$. 
For uncorrelated eigenvalues in the
complex plane, the Poisson distribution becomes \cite{Grob88}
\begin{equation}
  \label{Poisson}
  P_{\bar{\rm P}}(s)=\frac{\pi}{2}\,s\,e^{-\pi s^2/4}\:.
\end{equation}
This should not be confused with the Wigner distribution
for a Hermitian operator.~\cite{Hala95}

\begin{figure}
\begin{center}
\begin{tabular}{ccc}
  \hbox{\hspace{6.5mm} $\mu=0$}  &  \hbox{\hspace{2.75mm} $\mu=0.4$} &
  \hbox{\hspace{1.25mm} $\mu=3.0$ }\\[2mm]
  \epsfxsize=3.7cm\epsffile{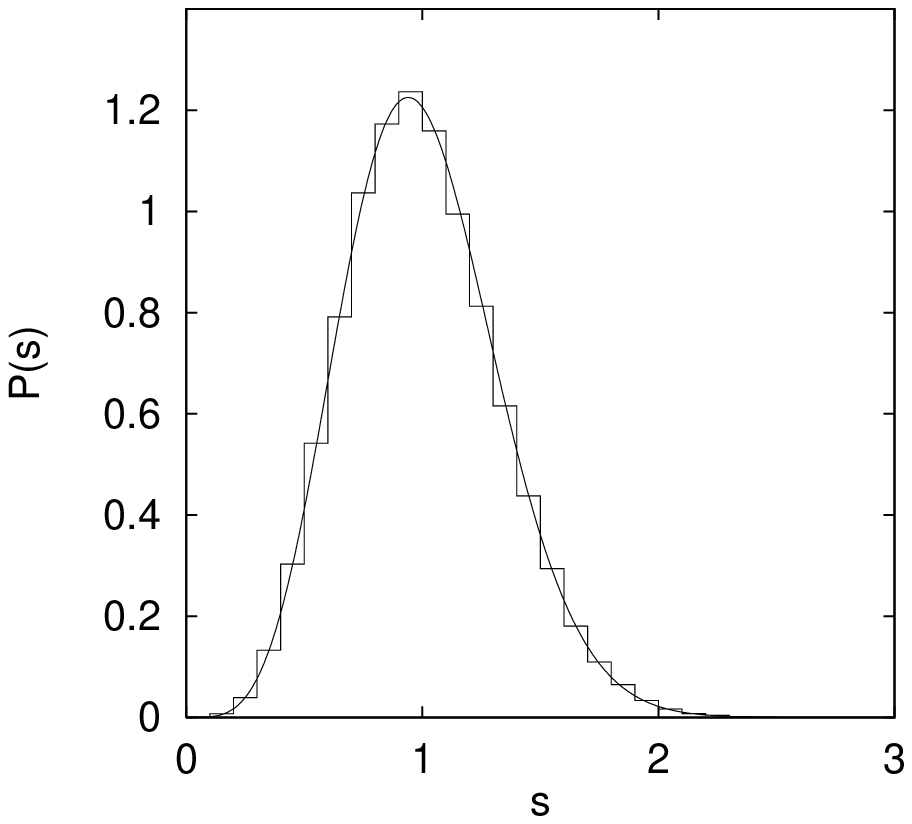} &
  \epsfxsize=3.7cm\epsffile{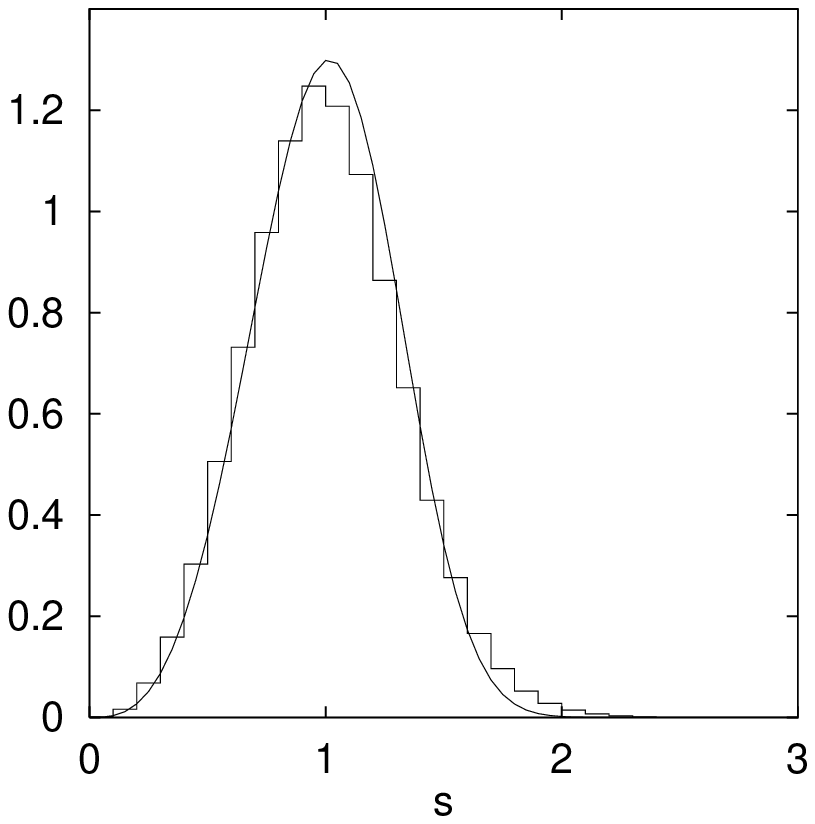} &
  \hbox{\hspace{-2.mm}\epsfxsize=3.7cm\epsffile{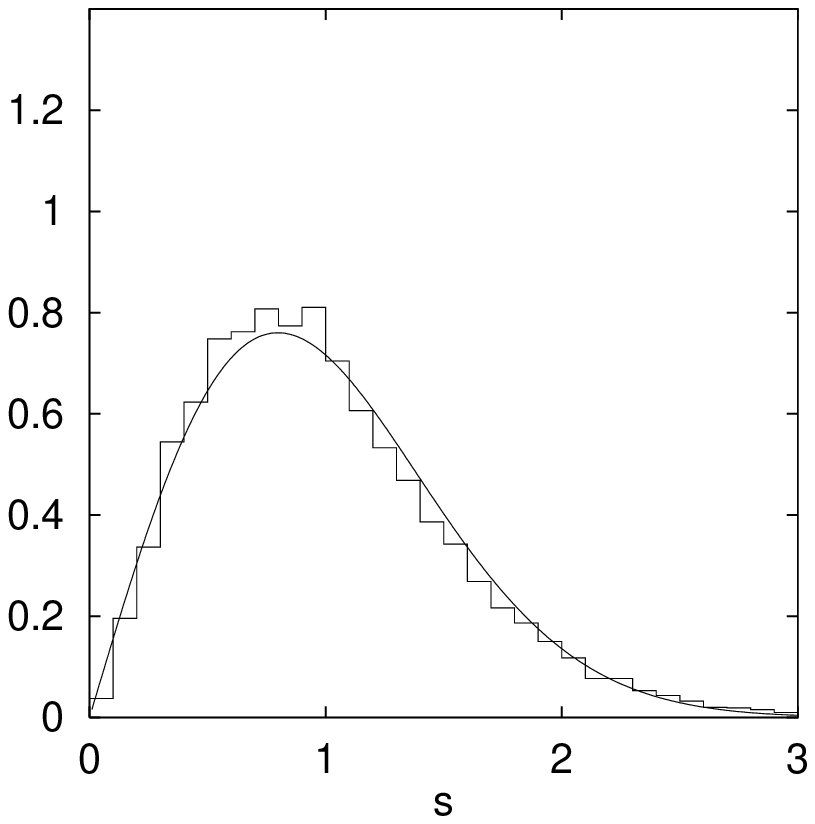}}
\end{tabular}
\end{center}
\vspace*{-3mm}
  \caption{Nearest-neighbor spacing distribution for two-color
    QCD with varying $\mu$.
    The analytic curves are the Wigner distribution, $P_{W} =
    262144/(729 \pi^3) s^4 \exp(-64/(9\pi)s^2)$ (left),
    the Ginibre distribution of Eq.~(\protect\ref{Ginibre}) (middle) and the
    Poisson distribution of Eq.~(\protect\ref{Poisson}) (right).}
  \vspace*{-3mm}
  \label{fig3}
\end{figure}

Our results for $P(s)$ are presented in Fig.~\ref{fig3}.  As a
function of $\mu$, we expect to find a transition from Wigner to
Ginibre behavior in $P(s)$. This was clearly seen in color-SU(3) with
$N_f=3$ flavors and quenched chemical potential,~\cite{Mark99} where
differences between both curves are more pronounced. For the
symplectic ensemble of color-SU(2) with staggered fermions, the Wigner
and Ginibre distributions are very close to each other and thus harder
to distinguish. They are reproduced by our preliminary data for
$\mu=0$ and $\mu=0.4$, respectively. Even in the deconfined phase,
where the effect of the chemical potential might order the system, 
the gauge fields  retain a considerable degree of randomness, which
apparently gives rise to quantum chaos.

For $\mu > 1.0$, the lattice results for $P(s)$ deviate substantially
from the Ginibre distribution and could be interpreted as Poisson
behavior, corresponding to uncorrelated eigenvalues. A plausible
explanation of the transition to Poisson behavior is provided by the
following two (related) observations. First, for large $\mu$ the terms
containing $e^\mu$ in Eq.~(\ref{Dirac}) dominate the Dirac matrix
giving
rise to uncorrelated eigenvalues. Second, for large $\mu$ the fermion density
on the finite lattice reaches saturation due to the limited box size and
the Pauli exclusion principle.

{\it Acknowledgments:}
This study was supported in part by FWF project P14435-TPH.
We thank the {\em ECT$^*$}, Trento, for hospitality during
various stages of this work.
We further thank B.A. Berg and T. Wettig for collaborations.~\cite{osaka}


\end{document}